# Periodic Motion near the Surface of Asteroids

Yu Jiang[1,2], Hexi Baoyin[1], Hengnian Li[2]

1. School of Aerospace Engineering, Tsinghua University, Beijing 100084, China
2. State Key Laboratory of Astronautic Dynamics, Xi'an Satellite Control Center, Xi'an 710043, China

Y. Jiang (✉) e-mail: jiangyu_xian_china@163.com (corresponding author)

**Abstract**. We are interested in the periodic motion and bifurcations near the surface of an asteroid. The gravity field of an irregular asteroid and the equation of motion of a particle near the surface of an asteroid are studied. The periodic motions around the major body of triple asteroid 216 Kleopatra and the OSIRIS-REx mission target-asteroid 101955 Bennu are discussed. We find that motion near the surface of an irregular asteroid is quite different from the motion near the surface of a homoplastically spheroidal celestial body. The periodic motions around the asteroid 101955 Bennu and 216 Kleopatra indicate that the geometrical shapes of the orbits are probably very sophisticated. There exist both stable periodic motions and unstable periodic motions near the surface of the same irregular asteroid. This periodic motion which is unstable can be resonant or non-resonant. The period-doubling bifurcation and pseudo period-doubling bifurcation of periodic orbits coexist in the same gravity field of the primary of the triple asteroid 216 Kleopatra. It is found that both of the period-doubling bifurcations of periodic orbits and pseudo period-doubling bifurcation of periodic orbits have four different paths. The pseudo period-doubling bifurcation found in the potential field of primary of triple asteroid 216 Kleopatra shows that there exist stable periodic orbits near the primary's equatorial plane, which gives an explanation for the motion stability of the triple asteroid 216 Kleopatra's two moonlets, Alexhelios and Cleoselene.

## 1. Introduction

Irregular asteroids are quite common in the solar system (Hartmann 2000). Space missions to asteroids (Emery et al. 2014) call to attention research about the laws of motion around irregular asteroids. In addition, several binary and triple asteroids have large size ratios (Marchis et al. 2010; Taylor and Margot 2011); research about the laws of motion around irregular asteroids are also useful for understanding the dynamical behaviors of moonlets of binary asteroids with large size ratios. The spacecraft or moonlet which is orbiting an irregular asteroid can be modeled by a massless particle (Scheeres 2012; Takahashi et al. 2013; Jiang et al. 2014; Jiang and Baoyin 2014; Chanut et al. 2014); if the particle is sufficiently far from the irregular asteroid, the asteroid can be approximately modelled as a sphere and the solar gravitation to the particle should be considered; conversely, if the particle is sufficiently close to the irregular asteroid, the irregular asteroidal gravity field provides the decisive effect on the motion of the particle, and the irregular asteroidal gravity field should be considered and the solar gravitation can be neglected (Borisov and Zakharov 2014; Wang et al. 2014).

Recently, a few irregular asteroids were selected to analyze motions around them, including asteroids 4 Vesta (Mondelo et al. 2010), 216 Kleopatra (Yu and Baoyin 2012; Jiang et al. 2014; Hirabayashi and Scheeres 2014), 433 Eros (Scheeres 2012), 1580 Betulia (Scheeres 2012), 1620 Geographos (Jiang et al. 2014), 4179 Toutatis (Scheeres 2012), 4769 Castalia (Takahashi et al. 2013; Jiang et al. 2014) and 6489 Golevka (Jiang et al. 2014). For the motion of a particle around an asteroid, the Jacobian integral and zero-velocity surfaces divide space into the forbidden region and the allowable region (Scheeres 2012; Yu and Baoyin 2012; Jiang et al. 2014). The velocity of the particle can be calculated by the effective potential of the asteroid (Jiang and Baoyin 2014). Stability of motion around equilibrium points is determined by the characteristics of the equilibrium points (Jiang et al. 2014). Jiang (2015) found that the topological classifications of equilibrium points and the topological classifications of periodic orbits around equilibrium points correspond to each other.

For some simple-shaped bodies, the existence, number, and stability of equilibrium points are discussed in detail; these simple-shaped bodies include a logarithm and a massive finite segment (Elipe and Riaguas 2003), a rotating homogeneous cube (Liu et al. 2011) and a dumbbell-shaped body (Li et al. 2013). Borisov and Zakharov (2014) discussed electrostatic charging and motion of dust particles around the surface of asteroids. Periodic orbits around simple-shaped bodies are also discussed; Riaguas et al. (1999) found several families of periodic orbits around a massive straight segment; in addition, Blesa (2006) calculated several families of periodic orbits around some simple planar plates. By using the direct geometrical look and a numerical method, periodic orbits near asteroid 216 Kleopatra (Yu and Baoyin 2012) and 4179 Toutatis (Scheeres 2012) can be classified into several families. These studies are about the motions in the potential of asteroids yet are not near the surface of the asteroid.

Liu et al. (2013) used a rotating cube to model the gravity field of an irregular asteroid, and discussed the surface equilibria and the motion around the surface equilibria. Tardivel et al. (2014) derived the impact equation of grains on the surface of an asteroid. Yu and Baoyin (2014) numerically calculated the hopping on the surface of asteroid 1620 Geographos. However, the motion near the surface differs from the remote motion and surface motion. The remote motion around an asteroid (Scheeres 2012) needs to consider the solar gravity, and the perturbation expansion with low-order (e.g. 10x10) Legendre coefficients can model the gravity field of an irregular asteroid. On the other hand, motion near the surface of an asteroid (Yu and Baoyin 2014) does not need to consider the solar gravity, just the high-precision gravity field of irregular asteroid, the coefficient of friction of the surface, and the mascons. The perturbation expansion with low Legendre coefficients is insufficient for the calculation of the motion near the surface of an asteroid, because the higher order terms take many iterations to converge, and even diverge at some points (Elipe and Riaguas 2003). A polyhedral model with sufficient vertices and faces (Werner and Scheeres 1997; Yu and Baoyin 2014) is necessary to calculate the gravity field and the irregular shape of asteroids.

With the goal of studying the periodic motion near the surface of an asteroid, we considered the irregular asteroidal gravity field, the equation of motion, and the effective potential. We discussed the periodic motion near the surfaces of asteroid 216 Kleopatra and asteroid 101955 Bennu; triple asteroid 216 Kleopatra has a large size ratio and asteroid 101955 Bennu is the target of the OSIRIS-REx mission. We find that the motion near the surface of an irregular asteroid is quite different from the motion near the surface of a homoplastic spheroidal celestial body, including whether a collision occurs in different frames as well as the complexity of the effective potential and periodic orbits. There exist orbits which are periodic in the body-fixed frame and will not collide with the body of the irregular asteroid; but they are non-periodic orbits in the inertial frame that will pass through the body of the irregular asteroid at the initial time. We found and analyzed four periodic orbits near the surface of asteroid 101955 Bennu and two periodic orbits near the surface of asteroid 216 Kleopatra. The geometric shapes of these orbits are very sophisticated. Stable and unstable periodic motions co-exist near the surface of the same irregular asteroid 101955 Bennu. We found two families of periodic orbits near the surface in the potential of the major body of the triple asteroid 216 Kleopatra. These are period-doubling bifurcation and pseudo period-doubling bifurcation.

**2. Equation of Motion and an Irregular Asteroid's Gravity Field**

Denote $\mathbf{r}$ as the body-fixed vector from the asteroid's mass center to the particle, $U(\mathbf{r})$ as the gravitational potential (Scheeres 2012) of the asteroid, $\boldsymbol{\omega}$ as the rotational angular velocity of the asteroid relative to the inertial space. To express all the vectors in the body-fixed frame, let

$$\boldsymbol{\omega} = \begin{pmatrix} \omega_x & \omega_x & \omega_x \end{pmatrix}^T, \quad \mathbf{v} = \dot{\mathbf{r}}, \quad \boldsymbol{\tau} = \dot{\boldsymbol{\omega}}, \tag{1}$$

and two $3 \times 3$ matrices

$$\hat{\boldsymbol{\omega}} = \begin{pmatrix} 0 & -\omega_z & \omega_y \\ \omega_z & 0 & -\omega_x \\ -\omega_y & \omega_x & 0 \end{pmatrix}, \quad \hat{\boldsymbol{\tau}} = \begin{pmatrix} 0 & -\tau_z & \tau_y \\ \tau_z & 0 & -\tau_x \\ -\tau_y & \tau_x & 0 \end{pmatrix}. \tag{2}$$

The effective potential is

$$V(\mathbf{r}) = -\frac{1}{2}(\boldsymbol{\omega} \times \mathbf{r}) \cdot (\boldsymbol{\omega} \times \mathbf{r}) + U(\mathbf{r}), \tag{3}$$

where the symbol $\times$ means exterior product while $\cdot$ means inner product.

Let $\mathbf{X} = \begin{bmatrix} \mathbf{r} \\ \mathbf{v} \end{bmatrix}$, $\mathbf{A} = \begin{pmatrix} \mathbf{I}_{3\times 3} & \mathbf{0}_{3\times 3} \\ -2\hat{\boldsymbol{\omega}} & -\hat{\boldsymbol{\tau}} \end{pmatrix}$, and $\mathbf{B}(\mathbf{X}) = \begin{pmatrix} \mathbf{0}_{3\times 3} \\ -\nabla V(\mathbf{r}) \end{pmatrix}$. Then, the dynamic equation (Jiang and Baoyin 2014) relative to the asteroid can be given as a first-order ordinary differential equation

$$\dot{\mathbf{X}} = \mathbf{A}\mathbf{X} + \mathbf{B}(\mathbf{X}). \tag{4}$$

Denote $\omega$ as the norm of the vector $\boldsymbol{\omega}$. Let the unit vector $\mathbf{e}_z$ of the body-fixed frame be expressed by $\boldsymbol{\omega} = \omega \mathbf{e}_z$, and let the asteroid rotate uniformly. Then the dynamic equations around the asteroid are

$$\begin{cases} \ddot{x} - 2\omega\dot{y} + \dfrac{\partial V}{\partial x} = 0 \\ \ddot{y} + 2\omega\dot{x} + \dfrac{\partial V}{\partial y} = 0, \\ \ddot{z} + \dfrac{\partial V}{\partial z} = 0 \end{cases} \quad (5)$$

where the effective potential $V = U - \dfrac{\omega^2}{2}(x^2 + y^2)$.

Using the effective potential, we can write the Jacobian integral as

$$H = \frac{1}{2}\left(\dot{x}^2 + \dot{y}^2 + \dot{z}^2\right) + V. \quad (6)$$

Usually, asteroids do have not enough mass to use their gravity to surmount the solid stress, so most asteroids have irregular shapes and an irregular gravity field (Ostro et al. 2000). Here we calculate the irregular gravity field of asteroid 216 Kleopatra. Asteroid 216 Kleopatra is a triple asteroid with two moonlets which are Alexhelios (S/2008 (216) 1) and Cleoselene (S/2008 (216) 2) (Descamps et al. 2011). The asteroid's gravitational potential (Werner and Scheeres 1997) is computed by

$$U = \frac{1}{2} G\sigma \sum_{e \in edges} \mathbf{r}_e \cdot \mathbf{E}_e \cdot \mathbf{r}_e \cdot L_e - \frac{1}{2} G\sigma \sum_{f \in faces} \mathbf{r}_f \cdot \mathbf{F}_f \cdot \mathbf{r}_f \cdot \omega_f, \quad (7)$$

and the gravitational force can be calculated by

$$\nabla U = -G\sigma \sum_{e \in edges} \mathbf{E}_e \cdot \mathbf{r}_e \cdot L_e + G\sigma \sum_{f \in faces} \mathbf{F}_f \cdot \mathbf{r}_f \cdot \omega_f, \quad (8)$$

where G=6.67×10$^{-11}$ m$^3$kg$^{-1}$s$^{-2}$ represents the gravitational constant, $\sigma$ represents the asteroid's bulk density; $\mathbf{r}_e$ and $\mathbf{r}_f$ are body-fixed vectors from field points to edge $e$ and face $f$, respectively. $\mathbf{E}_e$ and $\mathbf{F}_f$ are geometric parameters of edges and faces, respectively. $L_e$ is the integration factor between field points and edges $e$. $\omega_f$ is the signed angle relative to field points. The symbol $\cdot$ represents the scalar product.

The overall dimensions of 216 Kleopatra are $217 \times 94 \times 81$ km and estimated the rotational period is 5.385 h (Ostro et al. 2000); in addition, the estimated bulk density is 3.6 $g \cdot cm^{-3}$ (Descamps et al. 2011). The diameter of Alexhelios (S/2008 (216) 1) and Cleoselene (S/2008 (216) 2) were found to be $8.9 \pm 1.6$ km and $6.9 \pm 1.6$ km (Descamps et al. 2011), respectively. The physical model of 216 Kleopatra is calculated by a polyhedral model (Werner and Scheeres 1997) with 2048 vertices and 4096 faces which came from radar observations of asteroid 216 Kleopatra (Neese 2004). The Near-Earth Asteroid (NEA) 101955 Bennu (Campins et al. 2010) was discovered on September 11, 1999. A spacecraft based on the OSIRIS-REx mission will visit it; the mission will return samples to Earth for further detailed laboratory study (Emery et al. 2014). The mean diameter of asteroid 101955 Bennu is $492 \pm 20$ m (Nolan et al. 2013b), the sidereal rotation period is 4.288 h, and the bulk density is 0.95 $g \cdot cm^{-3}$ (Nolan et al. 2013a). Wang et al. (2014) calculated relative equilibrium points of 15 asteroids, 3 comets, and 5 satellites of planets; and found that only 2 asteroids have a number of relative equilibrium points larger than 5; they are asteroids 216 Kleopatra and 101955 Bennu, which have 7 and 9 relative equilibrium points, respectively. Figure 1 shows equilibrium points and the contour plot of the effective potential for asteroid 216 Kleopatra. From the diameters of these two asteroids, one can know that 216 Kleopatra has 4 relative equilibrium points outside and 3 relative equilibrium point inside the body while asteroid 101955 Bennu has 8 outside and 1 inside (Jiang et al. 2014; Wang et al. 2014; Jiang et al. 2015).

Equilibrium points satisfy $\nabla V = 0$. Thus the method to find out equilibrium points is (a) find a point $\mathbf{r}_0 = (x_0, y_0, z_0)$ and calculate the value of $\nabla V = \nabla U + \begin{bmatrix} -\omega^2 x & -\omega^2 y & 0 \end{bmatrix}^T$ at this point; (b) use the following formula to find the descent direction $\nabla(\nabla U) = G\sigma \sum_{e \in edges} \mathbf{E}_e \cdot \mathbf{L}_e - G\sigma \sum_{f \in faces} \mathbf{F}_f \cdot \boldsymbol{\omega}_f$; (c) in the descent direction, find a point $\mathbf{r}_1 = (x_1, y_1, z_1)$ which satisfies $|\nabla V(x_1, y_1, z_1)|$ is the smallest one; (d) repeat several steps in (b) and (c) such that $|\nabla V(x_1, y_1, z_1)|$ is sufficiently small. Equilibrium points inside the body of the asteroid are also calculated.

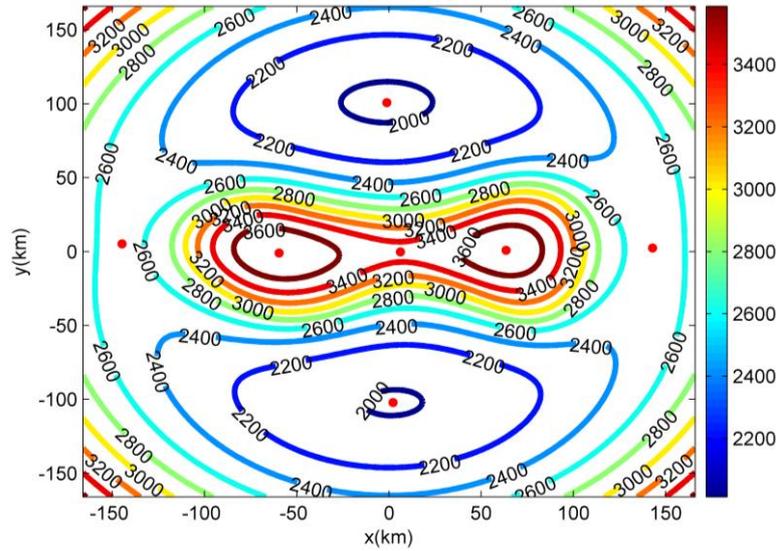

Fig. 1a  Seven equilibrium points and the contour plot of the effective potential for asteroid 216 Kleopatra, the unit of the effective potential is $m^2 \cdot s^{-2}$

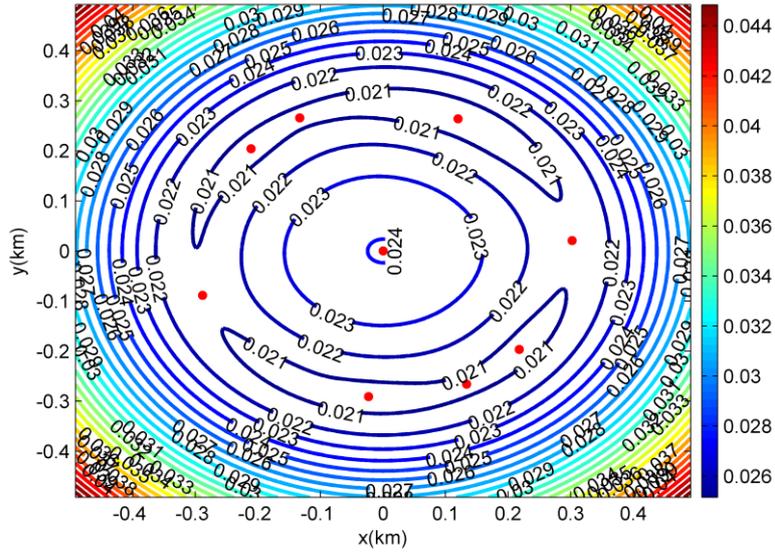

Fig. 1b    Nine equilibrium points and the contour plot of the effective potential for asteroid 101955 Bennu, the unit of the effective potential is $m^2 \cdot s^{-2}$

## 3. Stability and Resonance of Periodic Motion near the Surface of Asteroids

Motions of a particle near the surface of irregular asteroids are intricate and quite different from those near the surface of a homoplastic spheroidal celestial body. Throughout this paper, we calculate periodic orbits in the body-fixed frame, i.e. the rotating frame. In this section, we discuss the periodic motion near the surface of asteroids.

We write the dynamical equation as

$$\dot{\mathbf{X}} = \mathbf{f}(\mathbf{X}).  \qquad (9)$$

Let $S_p(T)$ be the set of periodic orbits which have period $T$. Taking a periodic orbit $p \in S_p(T)$, and using the $6 \times 6$ matrix $\nabla \mathbf{f} := \dfrac{\partial \mathbf{f}(\mathbf{z})}{\partial \mathbf{z}}$, we can write the state transition matrix (Hénon 1965; Scheeres 2012) for the periodic orbit as

$$\Phi(t) = \int_0^t \frac{\partial \mathbf{f}}{\partial \mathbf{z}}(p(\tau)) d\tau, \tag{10}$$

and the monodromy matrix for this periodic orbit $p \in S_p$ has the form

$$M = \Phi(T). \tag{11}$$

Eigenvalues of the monodromy matrix are Floquet multipliers of the periodic orbit.

**3.1 Periodic Motions near the surface in a Different Frame**

Using the hierarchical parameterization, we can calculate periodic orbits (Scheeres 2012; Yu and Baoyin 2012) by the grid search method. The grid search method uses the section plane to search the periodic orbits; the periodic orbit and the section plane are perpendicular. The periodic orbit is determined by five parameters: the azimuthal angle of the section plane $(\alpha, \beta)$, the position of the periodic orbit in the section plane (*u*, *v*) and the Jacobian constant *J*. Figure 2 shows a periodic orbit of the particle near the surface of asteroid 216 Kleopatra, the period is 10.049 h. Define the length unit for motion around 216 Kleopatra to be 219.0361 km, and the time unit to be 5.385 h. The initial position of this periodic orbit in the body-fixed frame is [1.172586 0.5120882 0.0891644], and the initial velocity in the body-fixed frame is [2.236803 -6.372973 0.3397389]. In Figure 2a, we plotted the body of asteroid 216 Kleopatra corresponding to its position at the initial time. From Figure 2, one can see that the orbit is periodic in the body-fixed frame, but in the inertial frame, it is a non-periodic orbit and passes through the body of asteroid 216 Kleopatra at the initial time. The mean radius of asteroid 216 Kleopatra is 135km (Descamps et al. 2011), the minimal distance between the orbit and the mass center of the asteroid is less than the mean radius of the asteroid. This periodic orbit indicates that the motion near the surface of irregular asteroids is quite different from the motion near the surface of a homoplastic spheroidal celestial body.

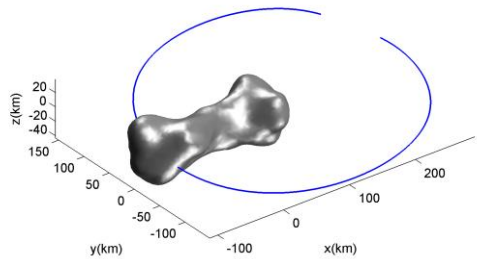 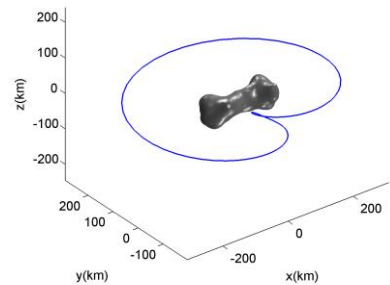

Figure 2a. A periodic orbit of the particle near the surface of the asteroid 216 Kleopatra in the inertial frame

Figure 2b. A periodic orbit of the particle near the surface of the asteroid 216 Kleopatra in the body-fixed frame

**3.2 Stability and Resonance**

Jiang (2015) discussed local periodic motion around equilibrium points, however, these local periodic motions around equilibrium points are only 1:1 resonant, which means the period ratio of the orbits and the asteroid's rotation period is 1:1. One can't find resonance periodic orbits around equilibrium points with other period ratios. Here we search for resonant motions which approach the surface of the irregular body with other period ratios. The relative equilibrium points influence relative periodic orbits (Jiang 2015), thus we choose asteroids 216 Kleopatra and 101955 Bennu to calculate periodic motions near their surfaces. Figure 3 shows four periodic orbits of the particle near the surface of asteroid 101955 Bennu and two periodic orbits near the surface of 216 Kleopatra. Define the length unit for the motion near the surface of 101955 Bennu to be 566.44 m, and the time unit to be 4.288 h. Define the length unit and the time unit for the motion near the surface of 216 Kleopatra to be 219.0361 km and 5.385 h, respectively. For a periodic orbit, the Floquet multipliers (that is, the eigenvalues of the monodromy matrix) are pairs of the form: $1$, $-1$, $\alpha^{\pm 1}$, $\cos\beta \pm i\sin\beta$, and $\sigma^{\pm 1}(\cos\tau \pm i\sin\tau)$, where $|\alpha| \in (0,1)$, $\beta \in (0,\pi)$, $\sigma > 0, \tau \in (0,\pi)$. Table 1 shows the initial positions and the initial velocities of six periodic orbits in the body-fixed frame. Periodic orbits 1-4 are relative to asteroid 101955 Bennu while periodic orbits 5-6 are relative to asteroid 216 Kleopatra. Table 2 presents Floquet multipliers of these periodic orbits near the surface of asteroid 101955 Bennu and 216 Kleopatra.

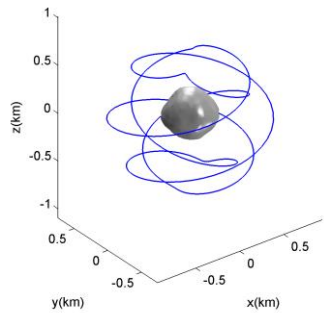

Fig. 3a Periodic orbit 1 near the surface of the asteroid 101955 Bennu

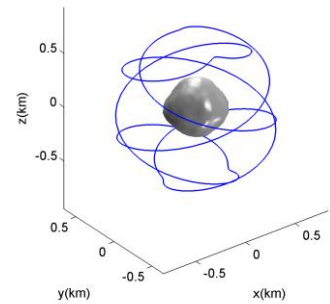

Fig. 3b Periodic orbit 2 near the surface of the asteroid 101955 Bennu

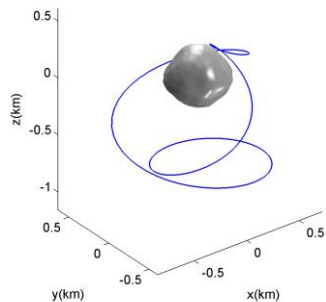

Fig. 3c Periodic orbit 3 near the surface of asteroid 101955 Bennu

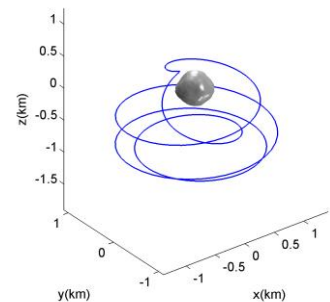

Fig. 3d Periodic orbit 4 near the surface of asteroid 101955 Bennu

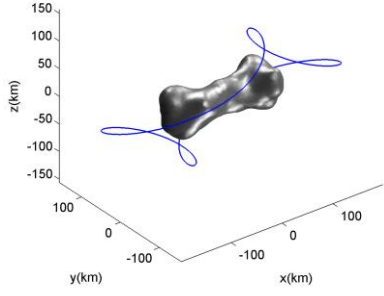
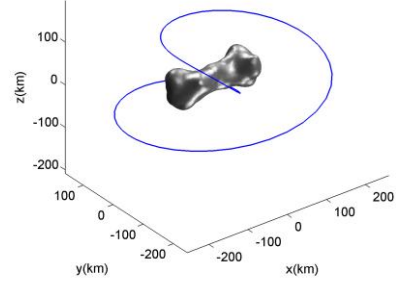

| Fig. 3e Periodic orbit 5 near the surface of asteroid 216 Kleopatra | Fig. 3f Periodic orbit 6 near the surface of asteroid 216 Kleopatra |
|---|---|

Table 1. The initial positions and the initial velocities of periodic orbits in the body-fixed frame, N: non-resonant

| Periodic Orbits | Positions | Velocities | Resonances |
|---|---|---|---|
| 1 | [-0.4622287   0.7255555   0.4877019] | [4.192610   1.411682   1.917582] | 7:1 |
| 2 | [0.7479705   0.9355176   0.05956148] | [5.320301   -4.069250   -1.877970] | 7:1 |
| 3 | [0.5153017   -0.8817847   -1.263795] | [-4.589355   -3.276604   0.7532408] | 3:1 |
| 4 | [1.940195   -0.3363006   -1.151279] | [-1.949430   -11.49121   -0.7283391] | 5:1 |
| 5 | [0.7950265   0.3624466   -0.1893687] | [2.573644   -3.896441   2.190076] | 2:1 |
| 6 | [0.3320805   0.1305879   -0.2952406] | [-0.6478920   -5.671883   1.156047] | N |

Table 2. Floquet multipliers of periodic orbits near the surfaces of asteroid 101955Bennu and 216 Kleopatra

| Orbits | $\lambda_1$, $\lambda_2$ | $\lambda_3$, $\lambda_4$ | $\lambda_5$, $\lambda_6$ |
|---|---|---|---|
| 1 | $1.374$, $\dfrac{1}{1.374}=0.726$ | $\cos 1.365° \pm i\sin 1.365°$ | 1, 1 |
| 2 | $\cos 9.770° \pm i\sin 9.770°$ | $\cos 1.202° \pm i\sin 1.202°$ | 1, 1 |
| 3 | $12.342$, $\dfrac{1}{12.342}=0.081$ | $1.167$, $\dfrac{1}{1.167}=0.857$ | 1, 1 |
| 4 | $1.226(\cos 14.188° \pm i\sin 14.188°)$ | $0.815(\cos 14.188° \pm i\sin 14.188°)$ | 1, 1 |
| 5 | $21.3005$, $\dfrac{1}{21.3005}=0.0469$ | $5.286$, $\dfrac{1}{5.286}=0.1892$ | 1, 1 |
| 6 | $8.5025$, $\dfrac{1}{8.5025}=0.1176$ | $-2.940$, $\dfrac{1}{-2.940}=-0.340$ | 1, 1 |

The period of periodic orbit 1 is 29.99 h, the ratio of the period of periodic orbit 1 and the asteroid's rotation period is 7:1; this periodic motion is resonant and unstable. The period is of periodic orbit 2 also 29.99 h; this periodic motion has the same period as periodic orbit 1. From Table 2, we can see that the Floquet multipliers of periodic orbits 1 and 2 are different; periodic orbit 2 is stable. These two periodic motions seem to be symmetrical. This result implies that periodic motions near the surface of an asteroid might have similar geometrical shapes and periods but have different topological cases and stability. Besides, the resonant motion near the surface of an asteroid may be stable. The period of periodic orbit 3 is 12.85 h; all Floquet multipliers of this periodic motion are real numbers, and this periodic motion is unstable. The period of periodic orbit 4 is 21.42 h; four Floquet multipliers of this periodic motion are complex numbers, and this periodic motion is also unstable. Periodic orbits 3 as well as 4 look asymmetric and are resonant. The ratio of the orbital periods and the asteroid's rotation period are 3:1 and 5:1, respectively. This result signifies that the unstable periodic orbits may have different distributions of Floquet multipliers.

The period of periodic orbit 5 is 10.636 h while the period of periodic orbit 6 is 7.0009 h. The ratio of the orbital periods and the asteroid's rotation period for periodic orbit 5 is 2:1. Periodic orbit 6 is a non-resonant orbit. Both of these two periodic orbits are unstable, as all the Floquet multipliers of these two periodic orbits are on the real axis. The Floquet multipliers of periodic orbit 5 are on the positive real axis while periodic orbit 6 has two Floquet multipliers on the negative real axis. To compare periodic orbits 3 and 5, one can see that the unstable periodic motion near the surface can be resonant or non-resonant.

## 4. Continuation of Periodic Orbit with Multi-parameters, Period-Doubling Bifurcations, and Pseudo Period-Doubling Bifurcations

As a periodic orbit continues (Hénon 1965; Tresaco et al. 2012), the topological cases of the periodic orbit may change, and period-doubling bifurcation may occur. The period-doubling bifurcation occurs if and only if 2 or 4 Floquet multipliers collide at -1 and leave the original region. The original region may be the x-axis or the unit cycle, and the other region omits the x-axis and the unit cycle.

The topological cases for the orbits with pure periodic cases contain seven different cases. These cases have at least two Floquet multipliers equal to 1. These seven cases are non-collisional and non-degenerate-real-saddle. The characteristics of these cases are presented in Table 3.

The topological cases for the periodic orbits with period-doubling contain four different cases. These cases have at least two Floquet multipliers equal to 1 and two equal to $-1$. The characteristics of these cases are presented in Table 4.

Table 3. The topological cases for the orbits with pure periodic cases

| Cases | The forms of the Floquet multipliers |
|---|---|
| 1 | $\gamma_j \left( \gamma_j = 1; j = 1, 2 \right)$ and $e^{\pm \sigma \pm i\tau} \left( \sigma, \tau \in \mathrm{R}; \sigma > 0, \tau \in (0, \pi) \right)$ |
| 2 | $\gamma_j \left( \gamma_j = 1; j = 1, 2 \right)$ and $e^{\pm i\beta_j} \left( \beta_j \in (0, \pi); j = 1, 2 \mid \beta_1 \neq \beta_2 \right)$ |
| 3 | $\gamma_j\left(\gamma_j=1; j=1,2\right)$ and $\operatorname{sgn}(\alpha_j)e^{\pm\alpha_j}\left(\alpha_j\in\mathrm{R},|\alpha_j|\in(0,1); j=1,2|\alpha_1\neq\alpha_2\right)$ |
| 4 | $\gamma_j\left(\gamma_j=1; j=1,2\right)$, $e^{\pm i\beta_j}\left(\beta_j\in(0,\pi), j=1\right)$ and $\operatorname{sgn}(\alpha_j)e^{\pm\alpha_j}\left(\alpha_j\in\mathrm{R},|\alpha_j|\in(0,1), j=1\right)$ |
| 5 | $\gamma_j\left(\gamma_j=1; j=1,2,3,4\right)$ and $e^{\pm i\beta_j}\left(\beta_j\in(0,\pi), j=1\right)$ |
| 6 | $\gamma_j\left(\gamma_j=1; j=1,2,3,4\right)$ and $\operatorname{sgn}(\alpha_j)e^{\pm\alpha_j}\left(\alpha_j\in\mathrm{R},|\alpha_j|\in(0,1), j=1\right)$ |
| 7 | $\gamma_j\left(\gamma_j=1; j=1,2,3,4,5,6\right)$ |

Table 4. The topological cases for the periodic orbits with period-doubling

| Cases | The forms of the Floquet multipliers |
|---|---|
| 8 | $\gamma_j\left(\gamma_j=1; j=1,2,3,4\right)$ and $\gamma_j\left(\gamma_j=-1; j=5,6\right)$ |
| 9 | $\gamma_j\left(\gamma_j=1; j=1,2\right)$ and $\gamma_j\left(\gamma_j=-1; j=3,4,5,6\right)$ |
| 10 | $\gamma_j\left(\gamma_j=-1; j=1,2\right)$, $\gamma_j\left(\gamma_j=1; j=3,4\right)$ and $e^{\pm i\beta_j}\left(\beta_j\in(0,\pi), j=1\right)$ |
| 11 | $\gamma_j\left(\gamma_j=-1; j=1,2\right)$, $\gamma_j\left(\gamma_j=1; j=3,4\right)$ and $\operatorname{sgn}(\alpha_j)e^{\pm\alpha_j}\left(\alpha_j\in\mathrm{R},|\alpha_j|\in(0,1), j=1\right)$ |

The period and Jacobian integral are taken as two parameters. If we only chose one parameter, such as the period, to continue the periodic orbit, then at a fold, the numerical continuation is impossible (Doedel et al. 2003); one can change another parameter when encountering this situation. Considering the topological cases and bifurcation of periodic orbit families, one can analyze the continuation of a periodic orbit with multiple parameters.

The continuation procedure (Muñoz-Almaraz et al. 2003) used is

$$\mathbf{X}_{i+1}=\mathbf{X}_i+\varepsilon\cdot\delta\mathbf{X}_i, \tag{12}$$

where

$$\delta\mathbf{X}_i=\begin{pmatrix}\boldsymbol{\omega}\times(\boldsymbol{\omega}\times\mathbf{r}_i)+\nabla U\\ \mathbf{v}_i\end{pmatrix}, \tag{13}$$

the subscript i means values of the i-th iteration, $\delta\mathbf{X}_i$ is the gradient direction, and the step size $\varepsilon$ is chosen such that $\mathbf{X}_{i+1}$ is the optimal value in the gradient direction. The numerical continuation is ended if the periodic orbit collides with the asteroid.

If one uses only one parameter to calculate the periodic orbit family, the continuation may stop when the parameter equals a local extreme point. The period-doubling bifurcations and the pseudo period-doubling bifurcations may occur when the periodic orbit family is continuing. The Floquet multipliers of periodic orbits will move while the continuation. If there exist at least two Floquet multipliers which collide at -1 and the topological cases of the periodic orbits after collision have no change, the pseudo period-doubling bifurcation occur; else if the topological cases of the periodic orbits after collision change, the period-doubling bifurcation occur. The sameness between the period-doubling bifurcation and the pseudo period-doubling bifurcation is they all have at least two Floquet multipliers collide at -1. The difference between them is whether the topological cases of the periodic orbits after collision change.

The period-doubling bifurcations for periodic orbit families have four different paths: **Period-doubling Bifurcation I**. The transfer of the topological case follows $\text{Case } 4 \rightarrow \text{Case } 11 \rightarrow \text{Case } 3$ or $\text{Case } 3 \rightarrow \text{Case } 11 \rightarrow \text{Case } 4$. **Period-doubling Bifurcation II**. The transfer of the topological case follows $\text{Case } 2 \rightarrow \text{Case } 10 \rightarrow \text{Case } 4$ or $\text{Case } 4 \rightarrow \text{Case } 10 \rightarrow \text{Case } 2$. **Period-doubling Bifurcation III**. The transfer of the topological case follows $\text{Case } 10 \rightarrow \text{Case } 9 \rightarrow \text{Case } 11$ or $\text{Case } 11 \rightarrow \text{Case } 9 \rightarrow \text{Case } 10$. **Period-doubling**

**Bifurcation IV**. The transfer of the topological case follows $\text{Case 5} \to \text{Case 8} \to \text{Case 6}$ or $\text{Case 6} \to \text{Case 8} \to \text{Case 5}$.

The pseudo period-doubling bifurcations for periodic orbit families have four different paths: **Pseudo Period-doubling Bifurcation I**. The transfer of the topological case follows $\text{Case 4} \to \text{Case 11} \to \text{Case 4}$; **Pseudo Period-doubling Bifurcation II**. The transfer of the topological case follows $\text{Case 2} \to \text{Case 10} \to \text{Case 2}$; **Pseudo Period-doubling Bifurcation III**. The transfer of the topological case follows $\text{Case 10} \to \text{Case 9} \to \text{Case 10}$; **Pseudo Period-doubling Bifurcation IV**. The transfer of the topological case follows $\text{Case 5} \to \text{Case 8} \to \text{Case 5}$.

The continuation phenomenon depends on the period-doubling bifurcations, pseudo period-doubling bifurcations, and the functional relation between the periodic orbit families, and the parameter. Now take the Jacobian integral as the parameter. We compute the periodic orbit family when the Jacobian integral is monotone increasing, and when it is monotone decreasing. Then we gets the following:

**Continuation phenomenon I**. There exists a local extreme point for the Jacobian integral, and no bifurcations. The numerical continuation stops. (The periodic orbit family cannot be extended at the local extreme point with the parameter-Jacobian integral; and there exists another parameter such that the periodic orbit family can be extended.)

**Continuation phenomenon II**. There exists a local extreme point for the Jacobian integral, and the period-doubling bifurcation occurs. There is another periodic orbit family with a double period.

**Continuation phenomenon III**. There exist period-doubling bifurcations with no local extreme point for the Jacobian integral, and there is another periodic orbit family with a double period.

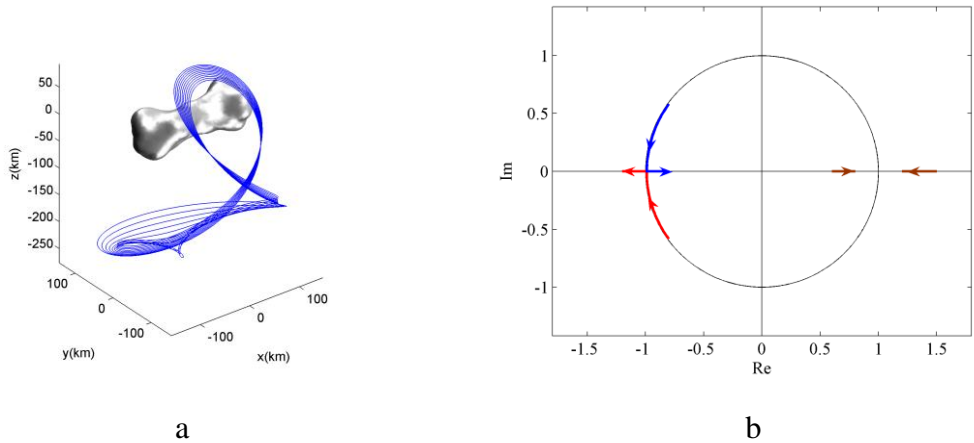

a                                       b

Fig. 4 Continuation of periodic orbit and period-doubling bifurcations in the potential of asteroid 216 Kleopatra

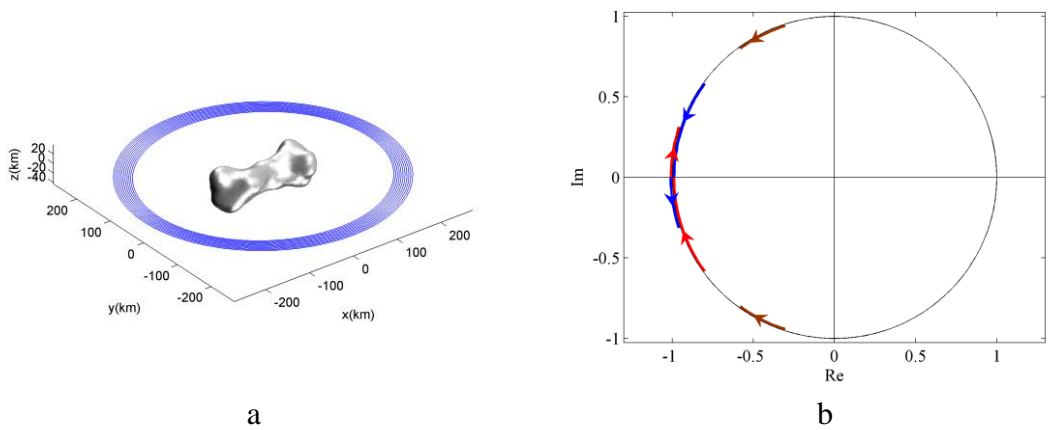

a                                       b

Fig. 5 Continuation of periodic orbit and pseudo period-doubling bifurcations in the potential of asteroid 216 Kleopatra

Asteroid 216 Kleopatra's shape is much more irregular than 101955 Bennu, and Kleopatra has two moonlets, which makes continuing the periodic orbit families around Kleopatra has more scientific significance than continuing them around Bennu.

So, we choose asteroid 216 Kleopatra to continue the periodic orbit families.

Figure 4a shows the continuation of the periodic orbit that has the period-doubling bifurcations. Figure 4b presents the motion of characteristic multipliers of periodic orbits when the numerical continuation of periodic orbit families is required, which leads to the transfer of topological cases of periodic orbits. From figure 4b, we can see that the period-doubling bifurcation occurs. The Jacobian integral varies between [-1160.454, -660.454] $m^2s^{-2}$. This family of periodic orbits is resonant. The period ratio of the periodic orbit and asteroid falls in the interval [2.0137, 2.0408], which implies that this periodic orbit family is resonant with the resonant ratio 2:1, however, the period ratio of the periodic orbit and asteroid for each of the periodic orbits in this family is not strictly 2:1. The bifurcation belongs to the period-doubling bifurcation I. The topological case of periodic orbits transitions from Case 4 to Case 11 and then to Case 3. The sojourn time at Case 11 is zero. In the continuation of this periodic orbit family, there is only one periodic orbit that has the topological Case 11. From this periodic orbit family, one can see that the resonant family of periodic orbits can contain period-doubling bifurcations. This family of periodic orbits cannot be continued by the period chosen to be the parameter.

Figure 5a shows the continuation of a periodic orbit that has the pseudo period-doubling bifurcations. Figure 5b presents the motion of characteristic multipliers of periodic orbits when at the numerical continuation of periodic orbit families, there is no transfer of topological cases of periodic orbits. From figure 5b, one can see that the pseudo period-doubling bifurcation occurs. The Jacobian integral varies in the range [-2553.908, -2453.908] $m^2s^{-2}$. This family of periodic orbits is non-resonant. The flight direction of the periodic orbit families is reversed, that means the rotation direction of 216 Kleopatra and the flight direction of the periodic orbit families are opposed. The periods of this family of periodic orbits vary over the interval [1.4110820, 1.5637778] $T_{216}$, where $T_{216}$ = 5.385 h is the rotation period of asteroid 216 Kleopatra. The pseudo period-doubling bifurcation belongs to the pseudo period-doubling bifurcation II. The topological case of periodic orbits follows the transition Case 2 → Case 10 → Case 2. The calculation of this periodic orbit family implies that there exist stable periodic orbits so that the Floquet multipliers collide at -1 and pass through each other. These two periodic orbit families in Figure 4a and 5a are different from periodic orbits shown in Figure 3 and cannot be connected to the periodic orbits shown in Figure 3. Periodic orbits in Figure 3 can also be used for the continuation, however, there are no period-doubling bifurcations or the pseudo period-doubling bifurcations occur in the continuation with periodic orbits in Figure 3.

Previous studies have continued the periodic orbit and found period-doubling bifurcations as well as pseudo period-doubling bifurcations of periodic orbit families in the gravitational field of asteroid 216 Kleopatra. This work is quite different from the continuation of the periodic orbit in the gravitational field of a simple-shaped body, such as a finite straight segment (Elipe and Lara 2003), a solid circular ring (Broucke and Elipe 2005), a homogeneous cube (Liu et al. 2011), and a massive annulus (Tresaco et al. 2012). The continuation of the periodic orbit in the gravitational field of a simple-shaped body can help one to understand the periodic orbit families, bifurcation of periodic orbit families, and stability of orbits around irregular minor celestial bodies. However, the simple-shaped body is not an irregular minor celestial body. Here triple asteroid 216 Kleopatra is taken as an example. The bigger moonlet, Alexhelios (S/2008 (216) 1), has a mass ratio of about 0.0003 relative to the major body. These two moonlets, Alexhelios (S/2008 (216) 1) and Cleoselene (S/2008 (216) 2) (Descamps et al. 2011), all have reversed nearly-circular quasi-periodic orbits relative to the primary of triple asteroid 216 Kleopatra. Figure 5 gives an explanation for the motion stability of these two moonlets. The periodic orbit family is reversed, nearly circular, and periodic, but is stable. During the continuation, the Floquet multipliers collide at -1. However, the collision of Floquet multipliers does not leads to the period-doubling bifurcation; after the collision, the Floquet multipliers pass through each other such that the periodic orbits are still stable. This implies that there exists a stable region in the potential field of the primary body of triple asteroid 216 Kleopatra, this stable region is consisted of periodic orbits or quasi-periodic orbits which are nearly circular, and the inclinations of the periodic orbits relative to the primary's body-fixed frame are approximately equal to zero.

## 5. Conclusions

We analyze the periodic motion near the surface of an asteroid. We discussed the irregular asteroidal gravity field, the equation of motion and the effective potential. Periodic motions near the surface of asteroid 216 Kleopatra and asteroid 101955

Bennu are investigated. The motion near the surface of an irregular asteroid is quite different from the motion near the surface of a homoplastic spheroidal celestial body; for example, there exist a periodic orbit such that the minimal distance between the orbit and the mass center of the asteroid is less than the mean radius of the asteroid. Moreover, periodic orbits near the surface of asteroid 101955 Bennu and 216 Kleopatra are analyzed. The result indicates that stable resonant periodic motions and unstable resonant periodic motions near the surface of the same irregular asteroid probably coexist.

We also investigated the continuation of a periodic orbit with multi-parameters. We found periodic orbit families in the potential of an asteroid to have four different kinds of period-doubling bifurcations and four kinds of pseudo period-doubling bifurcations. Numerical results illustrate that the period-doubling bifurcation and pseudo period-doubling bifurcation may coexist in the potential of an asteroid. We found a pseudo period-doubling bifurcation while a periodic orbit in the potential of the primary of triple asteroid 216 Kleopatra is continuing. The periodic orbit family is nearly circular and stable, and the orbit of the periodic orbit family and the equatorial plane of the primary are almost coincident, which implies that there exists a stable region in the potential of the primary of triple asteroid 216 Kleopatra, providing a point of view to understand the motion stability of the two moonlets.

**Acknowledgements**

This research was supported by the State Key Laboratory of Astronautic Dynamics Foundation (No. 2013ADL0202 & 2014ADL-DW0201), the National Basic Research Program of China (973 Program, 2012CB720004) and the National Natural Science